\documentclass[conference]{IEEEtran}
\IEEEoverridecommandlockouts
\usepackage{cite}
\usepackage{amsmath,amssymb,amsfonts}
\usepackage{algorithmic}
\usepackage{graphicx}
\usepackage{textcomp}
\usepackage{xcolor}
\def\BibTeX{{\rm B\kern-.05em{\sc i\kern-.025em b}\kern-.08em
    T\kern-.1667em\lower.7ex\hbox{E}\kern-.125emX}}

\usepackage[final]{microtype}
\usepackage[utf8]{inputenc}
\usepackage{multicol}
\usepackage{subfigure}
\usepackage{pifont}
\usepackage{enumitem}
\usepackage{multirow}

\usepackage[colorlinks,linkcolor=black,citecolor=blue,urlcolor=black]{hyperref}
\PassOptionsToPackage{pdfpagelabels=false}{hyperref} 

\usepackage{cleveref}
\crefformat{section}{\S#2#1#3}
\crefformat{subsection}{\S#2#1#3}
\crefformat{subsubsection}{\S#2#1#3}
\usepackage{titlesec}
\titlespacing\section{3pt}{3pt plus 2pt minus 2pt}{1pt plus 1pt minus 1pt}
\titlespacing\subsection{3pt}{2pt plus 2pt minus 1pt}{2pt plus 1pt minus 1pt}
\titlespacing\subsubsection{3pt}{2pt plus 2pt minus 1pt}{2pt plus 1pt minus 1pt}

\usepackage{fancyhdr}
\usepackage[normalem]{ulem}
\usepackage[final]{microtype}
\usepackage{fancyvrb}

\begin{document}

\title{\bf\LARGE FHEBench: Benchmarking Fully Homomorphic Encryption Schemes}

\author{
\IEEEauthorblockN{Lei Jiang}
\IEEEauthorblockA{Indiana University\\
jiang60@iu.edu}
\and
\IEEEauthorblockN{Lei Ju}
\IEEEauthorblockA{Shandong University\\
julei3@sdu.edu.cn}
}

\maketitle

\begin{abstract}
Fully Homomorphic Encryption (FHE) emerges one of the most promising solutions to privacy-preserving computing in an untrusted cloud. FHE can be implemented by various schemes, each of which has distinctive advantages, i.e., some are good at arithmetic operations, while others are efficient when implementing Boolean logic operations. Therefore, it is difficult for even cryptography experts let alone average users to choose the ``right'' FHE scheme to efficiently implement a specific application. Prior work only qualitatively compares few FHE schemes. In this paper, we present an empirical study, FHEBench, to quantitatively compare major FHE schemes.
\end{abstract}

\begin{IEEEkeywords}
Benchmarking, Fully Homomorphic Encryption
\end{IEEEkeywords}

\section{Introduction}
It is reluctant for average users to share sensitive information, e.g., healthcare records, to an untrusted cloud. New legislation such as EU GDPR or CCPA requires cloud computing enterprises to be more attentive about collecting, storing, utilizing, and transferring customer data.

\textit{Fully Homomorphic Encryption} (FHE)~\cite{Brakerski:TCT2014,Fan:CEAR2012,Cheon:ICTACIS2017,DUCAS:ICTACT2015,Chillotti:IACR2018} emerges as one of the most promising cryptographic technologies by allowing arbitrary computations to directly happen on encrypted data (ciphertexts). Particularly, compared to Multi-Party Computation, FHE does not require significant communication overhead between users and the cloud. FHE enables a user to encrypt their data and to send only ciphertexts to a cloud that can directly evaluate homomorphic operations on the ciphertexts. Final outputs are encrypted. Only the user who has the secret keys can decrypt outputs. In FHE-based cloud computing, users enjoy the end-to-end encryption. For example, FHE-enabled Machine Learning as a Service in the cloud performs neural network inferences~\cite{Dowlin:ICML2016} or trains neural networks~\cite{Lou:NIPS2020} on ciphertexts without secret keys. To guarantee the security, each homomorphic operation inevitably introduces a certain amount of noise into encrypted data. If there are too many homomorphic operations on the critical path, the accumulated noise may exceed a threshold, and the encrypted data cannot be decrypted correctly. FHE needs to periodically invoke a \textit{bootstrapping} operation to keep the noise in the encrypted data in check.

FHE computations can be implemented by various schemes, e.g., BGV~\cite{Brakerski:TCT2014}, BFV~\cite{Fan:CEAR2012}, CKKS~\cite{Cheon:ICTACIS2017}, FHEW~\cite{DUCAS:ICTACT2015}, and TFHE~\cite{Chillotti:IACR2018}. However, it is difficult for cryptography experts let alone average users to choose the ``right'' FHE scheme to efficiently implement a specific application. First, different FHE schemes have distinctive advantages. For instance, BFV is good at integer arithmetic, while TFHE can achieve fast Boolean algebra. Second, multiple FHE schemes can be used to implement the same type operation. Both BFV and BGV can efficiently support integer arithmetic, while both FHEW and TFHE can implement fast Boolean logic gates. Although prior works \cite{Melchor:ICSITC2018,Chaudhary:GUCON2019} qualitatively compare only few FHE schemes, there is totally a lack of a quantitative comparison of major FHE schemes.

In this paper, we first empirically compare state-of-the-art FHE libraries including Microsoft SEAL \cite{sealcrypto}, IBM HElib~\cite{Bergamaschi_HELIB2019}, CKKS~\cite{Cheon:HEAAN2019}, FHEW~\cite{Ducas:FHEW2019}, TFHE~\cite{Chillotti_TFHE2016}, and Palisade~\cite{PALISADE}. And then, we select competitive implementations among these libraries to characterize each FHE scheme. At last, we quantitatively compare the performance of various FHE schemes including BGV, BFV, CKKS, FHEW, and TFHE. Our study enables average users to choose the most efficient FHE scheme for their privacy-preserving applications.

\section{Background}
\label{s:back}

\subsection{FHE-based Cloud Computing}
\label{s:ppcc}

In FHE-based cloud computing, a client encrypts her data and sends only the ciphertext to the cloud. The untrusted server in the cloud performs FHE computations (e.g., neural network inference~\cite{Brutzkus:ICML2019} or training~\cite{Lou:NIPS2020}) on encrypted data, and sends the encrypted result back to the client. The end-to-end encryption of FHE-based cloud computing protects the client's privacy.

\subsection{Fully Homomorphic Encryption}

\textbf{FHE}. FHE allows computations to occur on ciphertexts. A plaintext $\mathbf{m}$ can be encrypted to a ciphertext $ct$ by an encryption function $enc$, so $ct=enc(\mathbf{m}, \mathbf{pk})$, where $\mathbf{pk}$ is the public key. $ct$ can be decrypted to $\mathbf{m}$ by a decryption function $dec$, i.e., $\mathbf{m}=dec(ct,\mathbf{sk})$, where $\mathbf{sk}$ is the secret key. An operation $\star$ is homomorphic if there is another operation $\bigcirc$ such that $enc(\mathbf{m_0},\mathbf{pk})\bigcirc enc(\mathbf{m_1},\mathbf{pk})=enc(\mathbf{m_0}\star \mathbf{m_1},\mathbf{pk})$, where $\mathbf{m_0}$ and $\mathbf{m_1}$ are plaintexts. To guarantee security, FHE introduces \textit{noise} into ciphertexts. An FHE operation enlarges the noise in the ciphertext which can accommodate only a limited number of FHE operations without a decryption failure. \textit{Bootstrapping} operations can keep the noise in check without decryption. However, the bootstrapping is extremely time-consuming. Most FHE libraries (e.g., Microsoft SEAL~\cite{sealcrypto}) do not even implement any bootstrapping. \textit{Level} FHE predefines a set of FHE parameters to allow a limited number of FHE multiplications (multiplicative depth) without bootstrapping.

\textbf{FHE Subroutines}. Major FHE schemes are built upon the variants of the Learning With Error (LWE) paradigm~\cite{Brakerski:TCT2014}. We use BFV~\cite{Fan:CEAR2012} as one example to explain FHE subroutines. We use $[r]$ to indicate a BFV ciphertext holding a message vector $r$, where the plaintext is a ring $\mathcal{R}_p = \mathbb{Z}_p[x]/(x^n+1)$ with \textit{plaintext modulus} $p$ and \textit{cyclotomic order} $n$. A ciphertext $[r]\in\mathcal{R}_q^2$ is a set of two polynomials in a quotient ring $\mathcal{R}_q$ with \textit{ciphertext modulus} $q$. BFV operations include:
\begin{itemize}[nosep,leftmargin=*]

\item \textbf{Step($\mathbf{\lambda}$)}. For a security parameter $\lambda$, we set a ring size $n$, a ciphertext modulus $q$, a special modulus $p$ coprime to $q$, a key distribution $\chi$, and an error distribution $\Omega$ over $R$.

\item \textbf{Encryption}. A BFV ciphertext is \textit{encrypted} as a vector of two polynomials $(c_0,c_1)\in\mathcal{R}_q^2$. Specifically, we have $c_0=-a$ and $c_1=a\cdot s+\frac{q}{p}m+e_0$, where $a$ is a uniformly sampled polynomial. $s$ and $e_0$ are polynomials whose coefficients drawn from $\mathcal{X}_{\sigma}$ ($\sigma$ is the standard deviation). 

\item \textbf{Decryption}. The \textit{decryption} computes $\frac{p}{q}(c_0 \cdot s+c_1)=m+\frac{p}{q}e_0$. When $\frac{q}{p}\gg e_0$, $e_0$ can be removed. Each FHE operation enlarges $e_0$. To successfully decrypt the ciphertext, a bootstrapping is required to keep $e_0$ in check. Or we have to choose a larger $q$ to accommodate a larger depth of FHE operations, since a larger $q$ can omit a larger $e_0$. 

\item \textbf{Addition}. For two ciphertexts $c_0=(c_{0,0},c_{1,0})$ and $c_1=(c_{0,1},c_{1,1})$, a FHE addition simply adds the polynomials of the two ciphertexts and outputs the result ciphertext $c=(c_{0,0}+c_{0,1}, c_{1,0}+c_{1,1})$.

\item \textbf{Multiplication}. During a FHE multiplication, the polynomials are first lifted to a large modulus $Q$. Multiple polynomial additions and multiplications are performed for the FHE multiplication with $Q$. And then, the polynomials are scaled back to the modulus $q$ from $Q$. A FHE multiplication adds another polynomial in the resulting ciphertext, so a \textit{relinearization} operation is required to transform the ciphertext back to a pair of polynomials $(c_0,c_1)$.

\item \textbf{Rotation}. BFV supports the SIMD-style batching~\cite{Fan:CEAR2012}, which can pack $m$ plaintexts $[p_0,\ldots,p_{m-1}]$ into a ciphertext, e.g., $(c_0,c_1)$. A homomorphic multiplication or addition happening between two ciphertexts $(c_{0,0},c_{1,0})$ and $(c_{0,1},c_{1,1})$ is equivalent to $m$ multiplications or additions between $[p_{0,0},\ldots,p_{m-1,0}]$ and $[p_{0,1},\ldots,p_{m-1,1}]$. For instance, a FHE addition can achieve $p_{0,0}+p_{0,1}$, $\ldots$, $p_{m-1,0}+p_{m-1,1}$. A FHE rotation operation rotates a ciphertext by several plaintext slots. For instance, $[p_{0,0},\ldots,p_{m-1,0}]$ can be rotated to $[p_{n,0},\ldots, p_{m-1,0}, p_{0,0},\ldots,p_{n-1,0}]$ by $n$ plaintext slots. And then, another FHE addition can obtain $p_{n,0}+p_{0,1}$, $\ldots$, $p_{n-1,0}+p_{m-1,1}$.
\end{itemize}
Some FHE schemes including BGV~\cite{Halevi:ICTACT2015} and CKKS~\cite{Cheon:ICTACIS2017} have similar subroutines to BFV, but each of them has their own unique implementation details. Other FHE schemes such as FHEW~\cite{DUCAS:ICTACT2015}, and TFHE~\cite{Chillotti:IACR2018} support homomorphic Boolean logic operations, e.g., XOR, AND, and NAND.

\begin{table}[t!]
\footnotesize
\setlength{\tabcolsep}{3pt}
\centering
\caption{The comparison of FHE schemes.}
\vspace{-0.1in}
\begin{tabular}{|c||c|c|c|c|c|}
\hline
Operation                  & BFV & BGV & CKKS & FHEW & TFHE \\\hline\hline
Native Add/Sub             & \ding{51} & \ding{51} & \ding{51} & \ding{55} & \ding{55} \\\hline
Native Mult                & \ding{51} & \ding{51} & \ding{51} & \ding{55} & \ding{55} \\\hline
SIMD                       & \ding{51} & \ding{51} & \ding{51} & \ding{51} & \ding{51} \\\hline
Boolean Logic              & \ding{55} & \ding{51} & \ding{55} & \ding{51} & \ding{51} \\\hline
$<1s$ Bootstrapping        & \ding{55} & \ding{55} & \ding{55} & \ding{51} & \ding{51} \\\hline
\end{tabular}
\label{t:he_op_comp}
\vspace{-0.2in}
\end{table}

\subsection{FHE Schemes}

FHE can be implemented by various HE schemes including BFV \cite{Fan:CEAR2012}, BGV~\cite{Halevi:ICTACT2015}, CKKS~\cite{Cheon:ICTACIS2017}, FHEW~\cite{DUCAS:ICTACT2015}, and TFHE~\cite{Chillotti:IACR2018}. Table~\ref{t:he_op_comp} compares different features of various FHE schemes. BGV, BFV, CKKS are based on the Ring LWE (RLWE) paradigm~\cite{Brakerski:TCT2014,Fan:CEAR2012}. They heavily rely on Number Theoretic Transform (NTT) and modular arithmetic operations. The RLWE schemes are good at computing vectorial multiplications and additions, and support the SIMD batching, which allows a vector of plaintexts to be encrypted as a single ciphertext. BGV and BFV support only integer arithmetic operations, while CKKS~\cite{Cheon:ICTACIS2017} is more efficient when processing complex numbers or fixed-point computations. Compared to BGV, BFV uses a different version of relinearization in a FHE multiplication. And CKKS supports only approximate computing, where each CKKS operation slightly modifies the value of the fraction part of the ciphertext. The bootstrapping operations of BGV, BFV and CKKS are time-consuming. For instance, the latency of a BGV bootstrapping operation costs several hundred seconds~\cite{Halevi:ICTACT2015}. In contrast, FHEW and TFHE are based on the LWE paradigm~\cite{Chillotti:IACR2018,DUCAS:ICTACT2015}. They are more powerful when processing binary logic operations, e.g., NAND, AND, XOR, and OR. By homomorphic binary logic operations, they can implement not only linear homomorphic operations, e.g., additions and multiplications, but also non-linear homomorphic operations, e.g., ReLU activations. Moreover, they have to perform a fast bootstrapping operation at the end of each FHE logic operation. A TFHE bootstrapping requires only $13ms$ on a CPU. FHEW and TFHE frequently invoke fast Fourier transform (FFT). 

\begin{table}[t!]
\footnotesize
\setlength{\tabcolsep}{3pt}
\centering
\caption{The comparison of FHE libraries.}
\vspace{-0.1in}
\begin{tabular}{|c||c|c|c|c|c|c|}
\hline
\multirow{3}{*}{Scheme} & \multicolumn{6}{c|}{HE Libraries}                                             \\\cline{2-7}
                        & SEAL & HElib & CKKS & FHEW & TFHE   & Palisade        \\
												& \cite{sealcrypto} 3.7.2 & \cite{Bergamaschi_HELIB2019} 2.2.1 & \cite{Cheon:HEAAN2019} 2.1 & \cite{Ducas:FHEW2019} 2  & \cite{Chillotti_TFHE2016} 1.0.1 & \cite{PALISADE} 1.11.5   \\\hline\hline
BFV                     & \ding{51}  & \ding{55}  & \ding{55} & \ding{55} & \ding{55} & \ding{51}       \\\hline
BGV                     & \ding{55}  & \ding{51}  & \ding{55} & \ding{55} & \ding{55} & \ding{51}       \\\hline
CKKS                    & \ding{51}  & \ding{51}  & \ding{51} & \ding{55} & \ding{55} & \ding{51}       \\\hline
FHEW                    & \ding{55}  & \ding{55}  & \ding{55} & \ding{51} & \ding{55} & \ding{51}       \\\hline
TFHE                    & \ding{55}  & \ding{55}  & \ding{55} & \ding{55} & \ding{51} & \ding{51}       \\\hline
\end{tabular}
\label{t:he_lib_comp}
\vspace{-0.2in}
\end{table}

\subsection{FHE Libraries}

Multiple libraries are built to implement FHE schemes by both academia and industry. For instance, Microsoft SEAL~\cite{sealcrypto} implements BFV and CKKS, while Palisade~\cite{PALISADE} supports all major FHE schemes. The comparison between various FHE libraries is shown in Table~\ref{t:he_lib_comp}. Different FHE libraries have many distinctive implementation details that have a huge impact on the performance of FHE operations. For instance, SEAL and Palisade use different methods to implement the double Chinese Remainder Theorem (CRT) Residue Number System (RNS) of BFV, thereby requiring different latencies for the same FHE operations.

\begin{figure*}[t!]
\centering
\subfigure[Key Generation]{
   \includegraphics[width=1.3in]{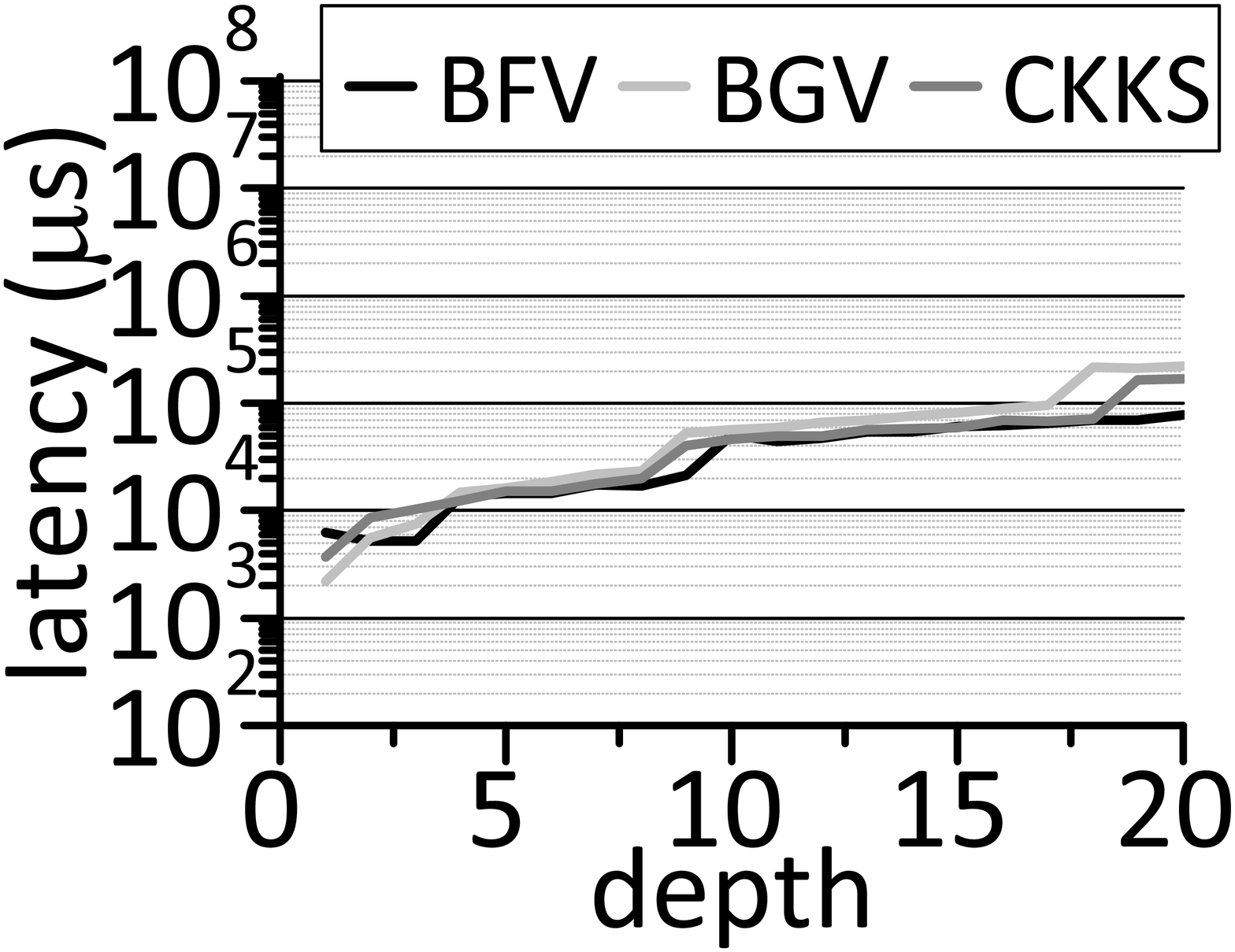}
   \label{f:ben_he_keygen}
}
\hspace{-0.15in}
\subfigure[Relinearization Key]{
   \includegraphics[width=1.3in]{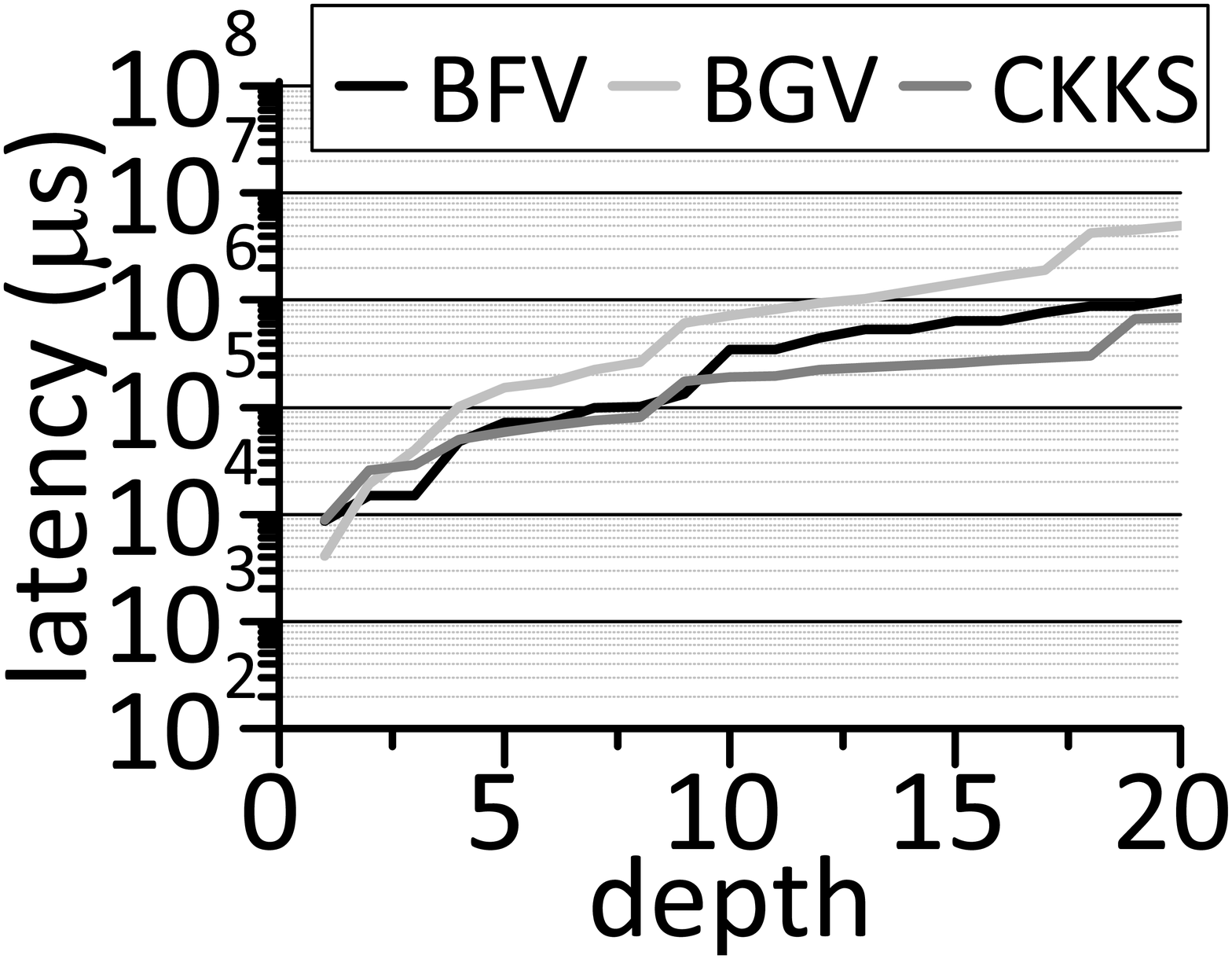}
   \label{f:ben_he_keyrelin}
}
\hspace{-0.15in}
\subfigure[Rotation Key]{
   \includegraphics[width=1.3in]{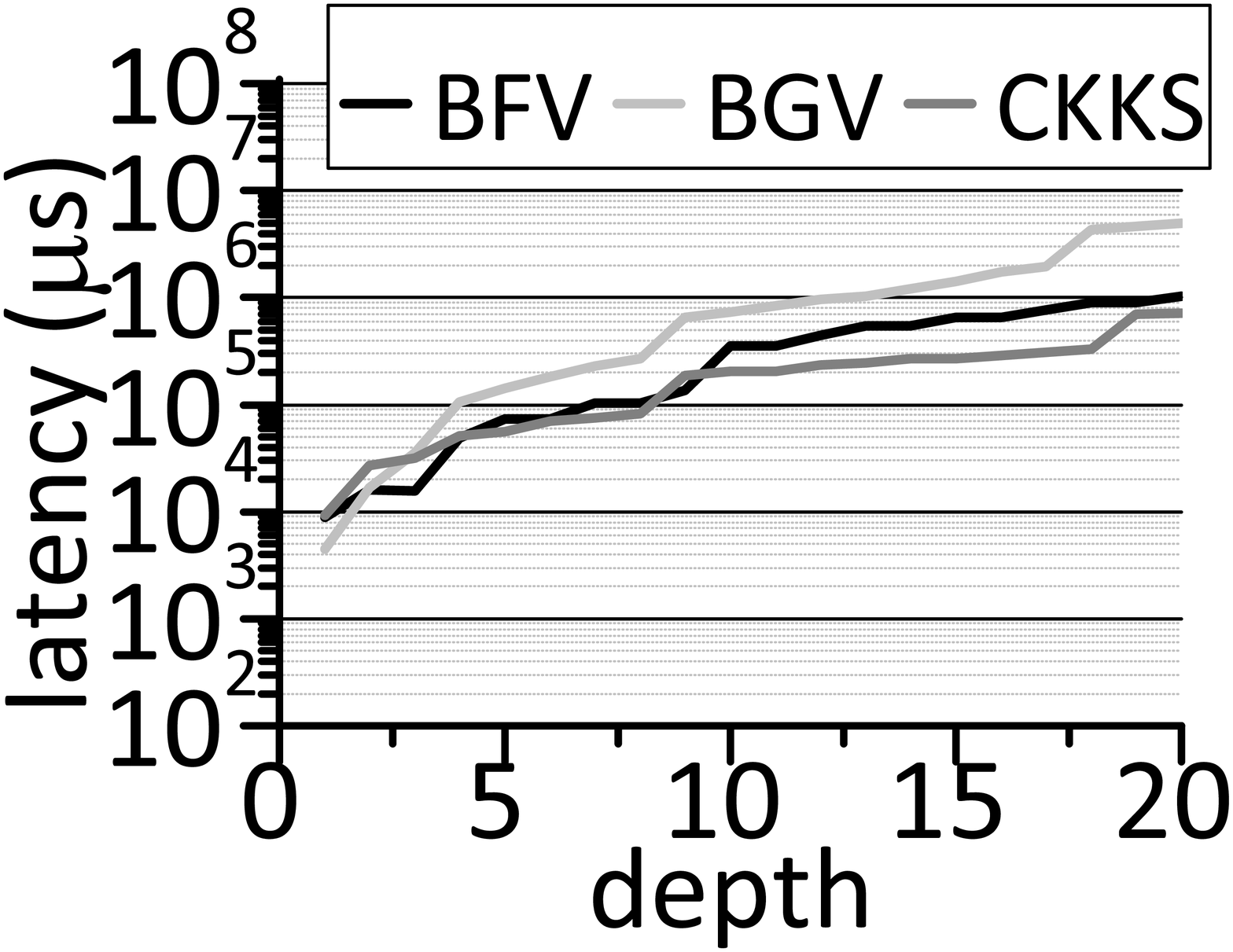}
   \label{f:ben_he_keyrot}
}
\hspace{-0.15in}
\subfigure[Encryption]{
   \includegraphics[width=1.3in]{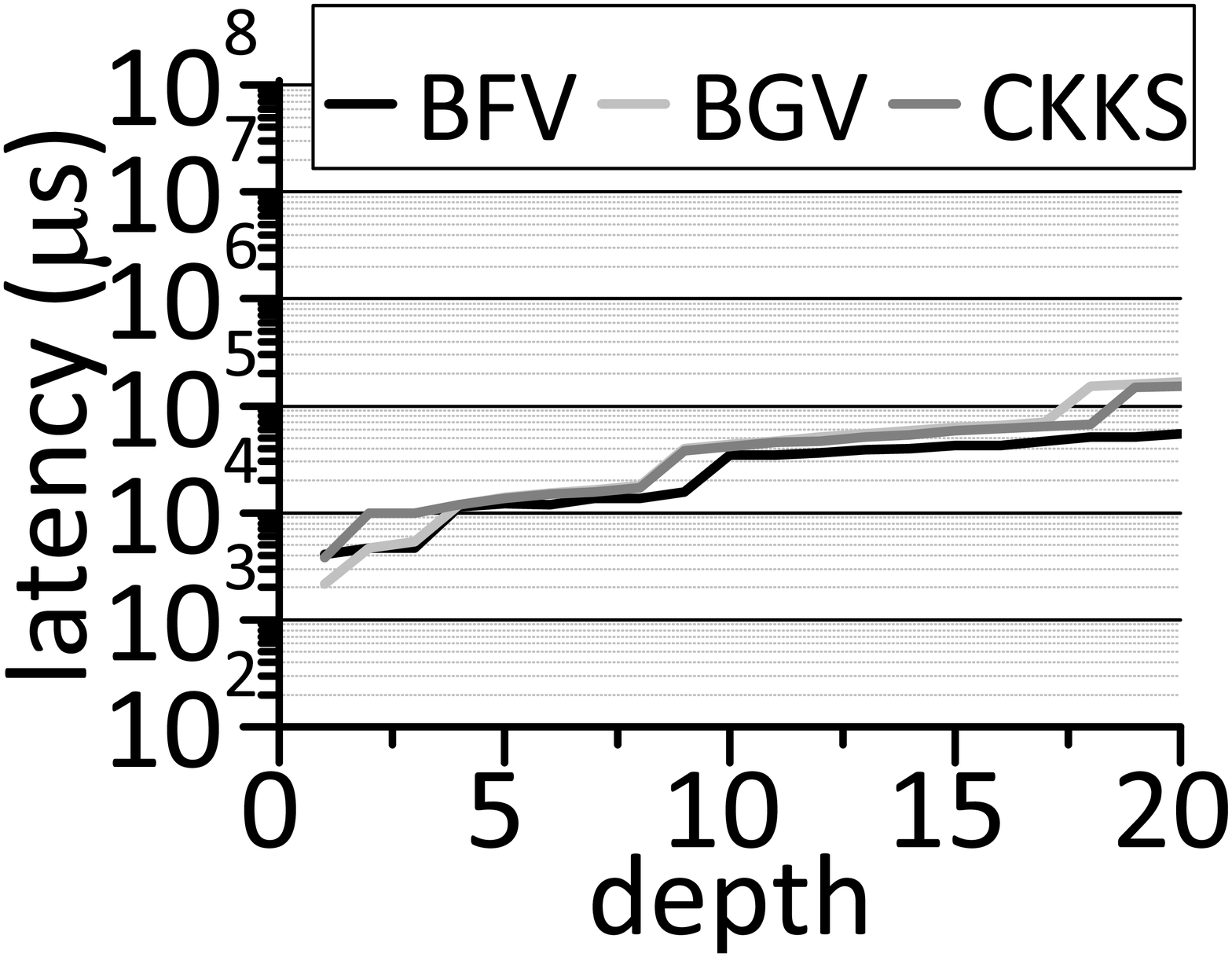}
   \label{f:ben_he_encrypt}
}
\hspace{-0.15in}
\subfigure[Decryption]{
   \includegraphics[width=1.3in]{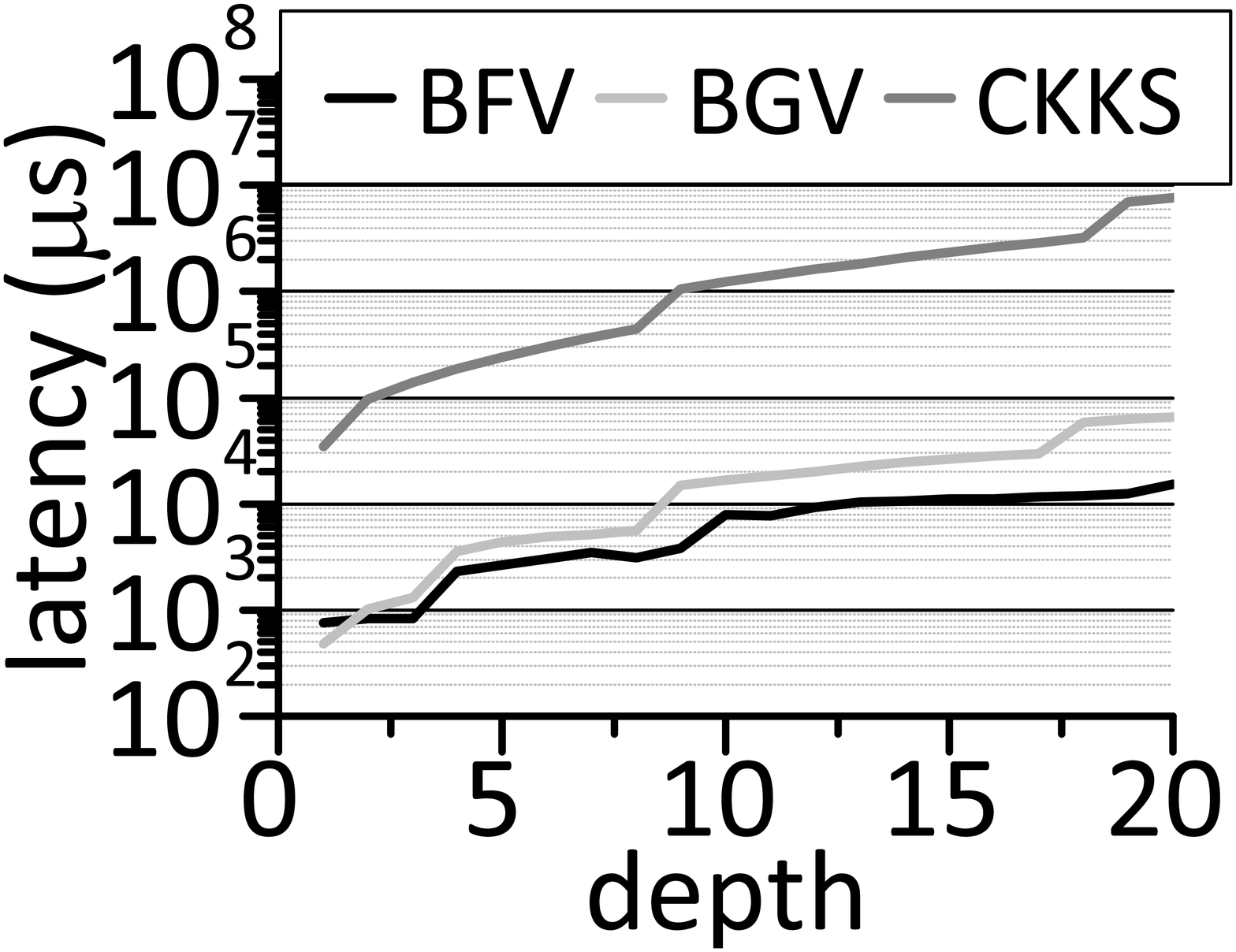}
   \label{f:ben_he_decrypt}
}
\vspace{-0.1in}
\caption{The latency comparison of client setup operations for FHE between various schemes.}
\label{f:ben_peri_comparison}
\vspace{-0.05in}
\end{figure*}

\begin{figure*}[t!]
\vspace{-0.05in}
\centering
\subfigure[AddCC]{
   \includegraphics[width=1.3in]{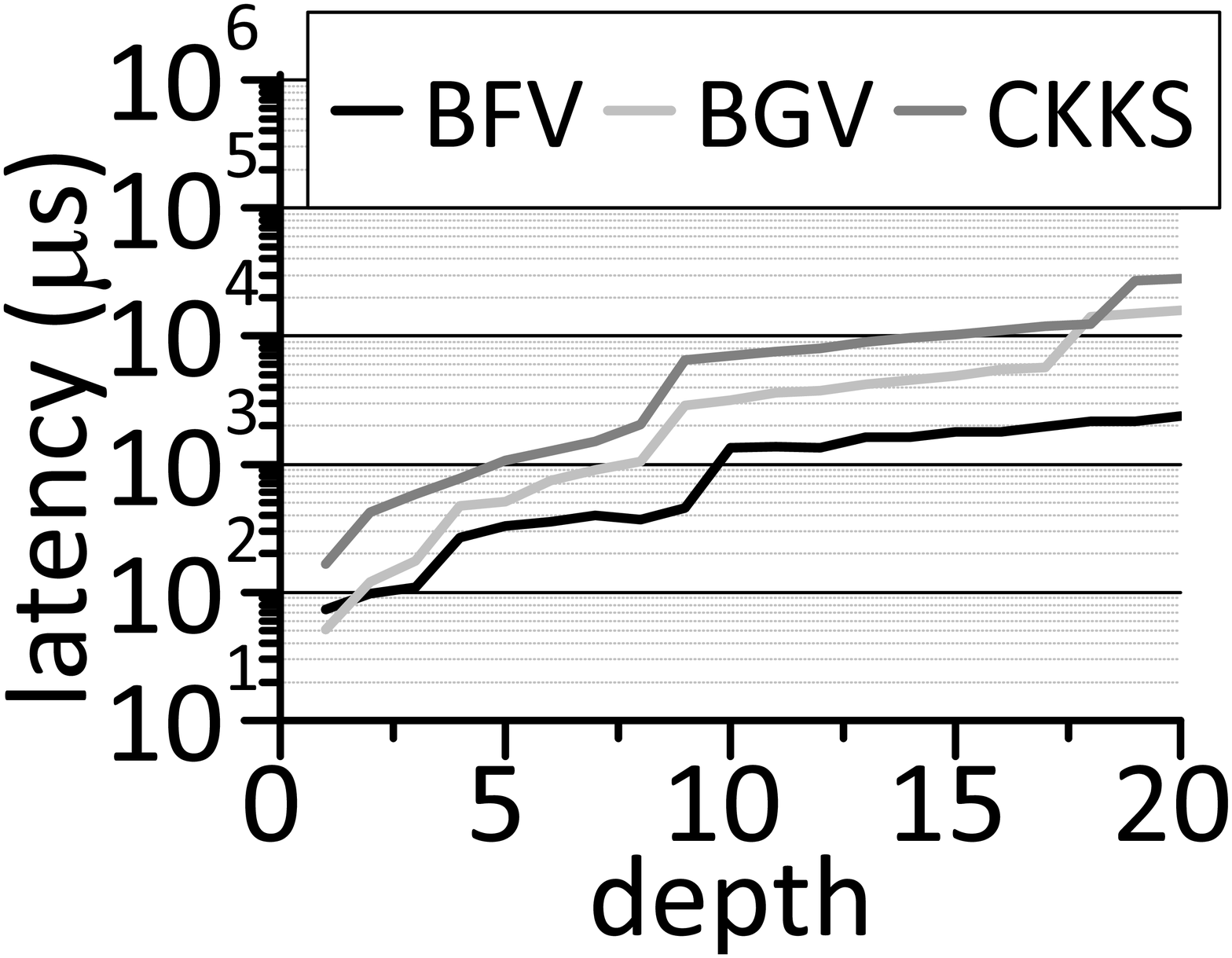}
   \label{f:ben_he_addcc}
}
\hspace{-0.15in}
\subfigure[AddCP]{
   \includegraphics[width=1.3in]{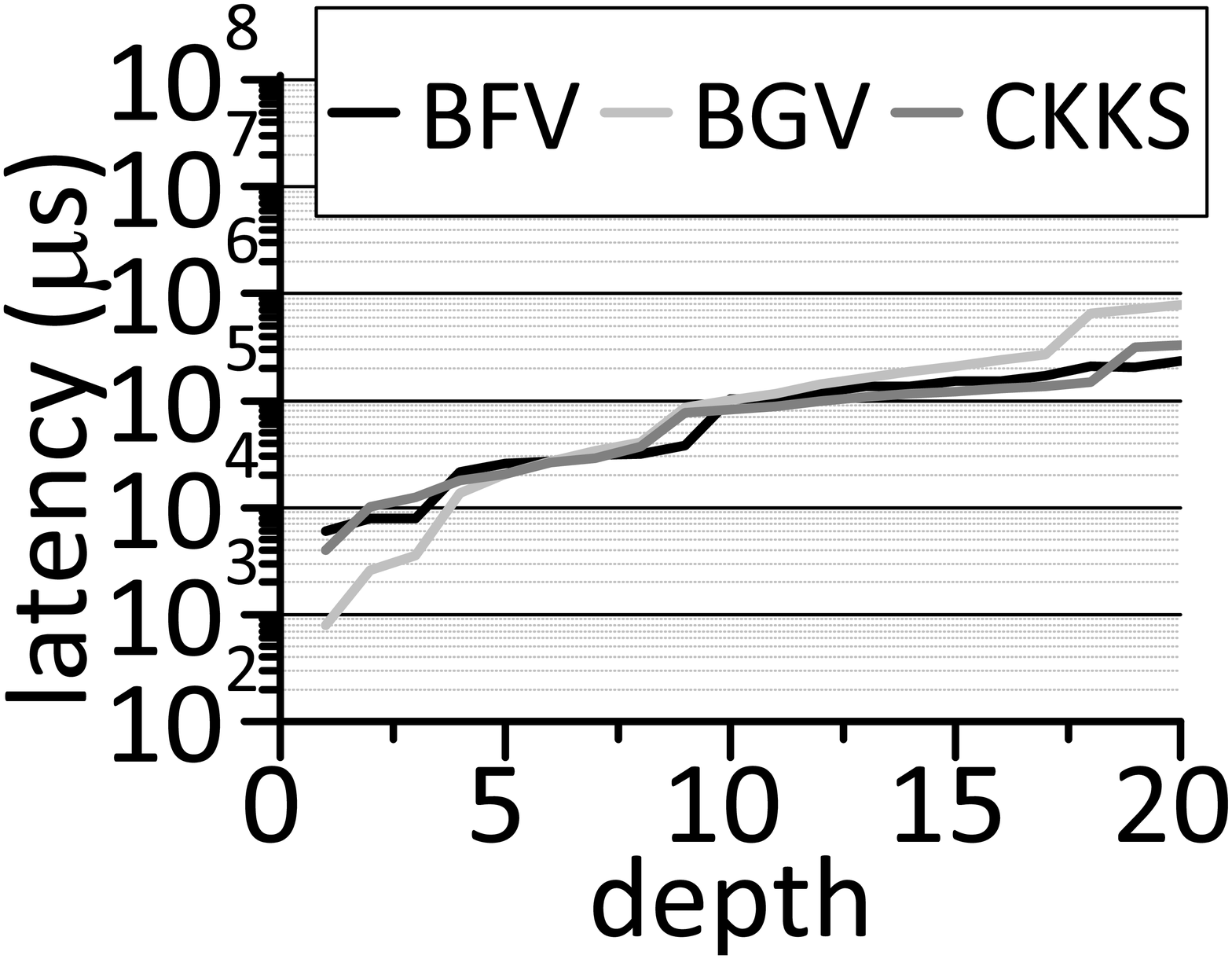}
   \label{f:ben_he_addcp}
}
\hspace{-0.15in}
\subfigure[MultCC]{
   \includegraphics[width=1.3in]{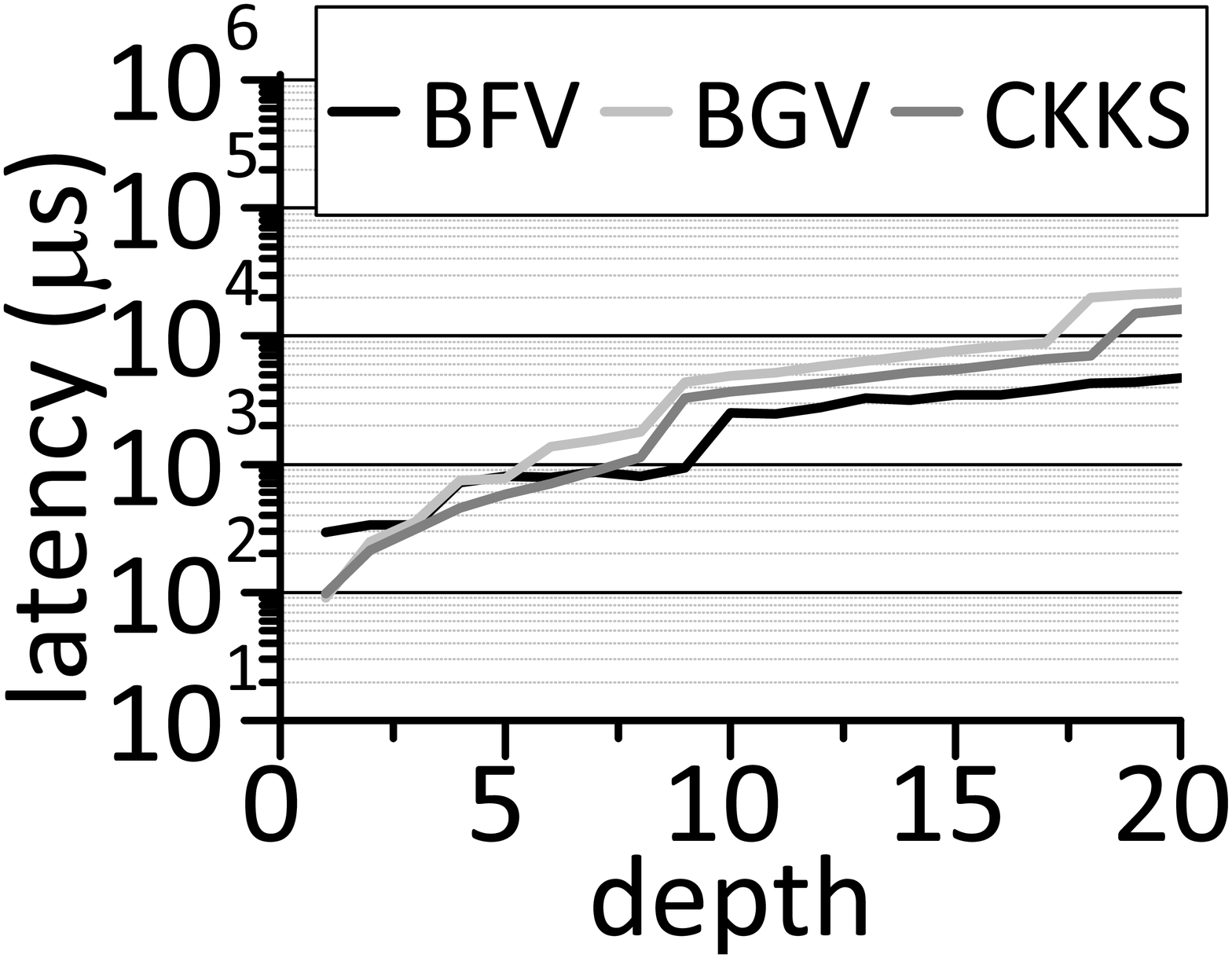}
   \label{f:ben_he_mulcc}
}
\hspace{-0.15in}
\subfigure[MultCP]{
   \includegraphics[width=1.3in]{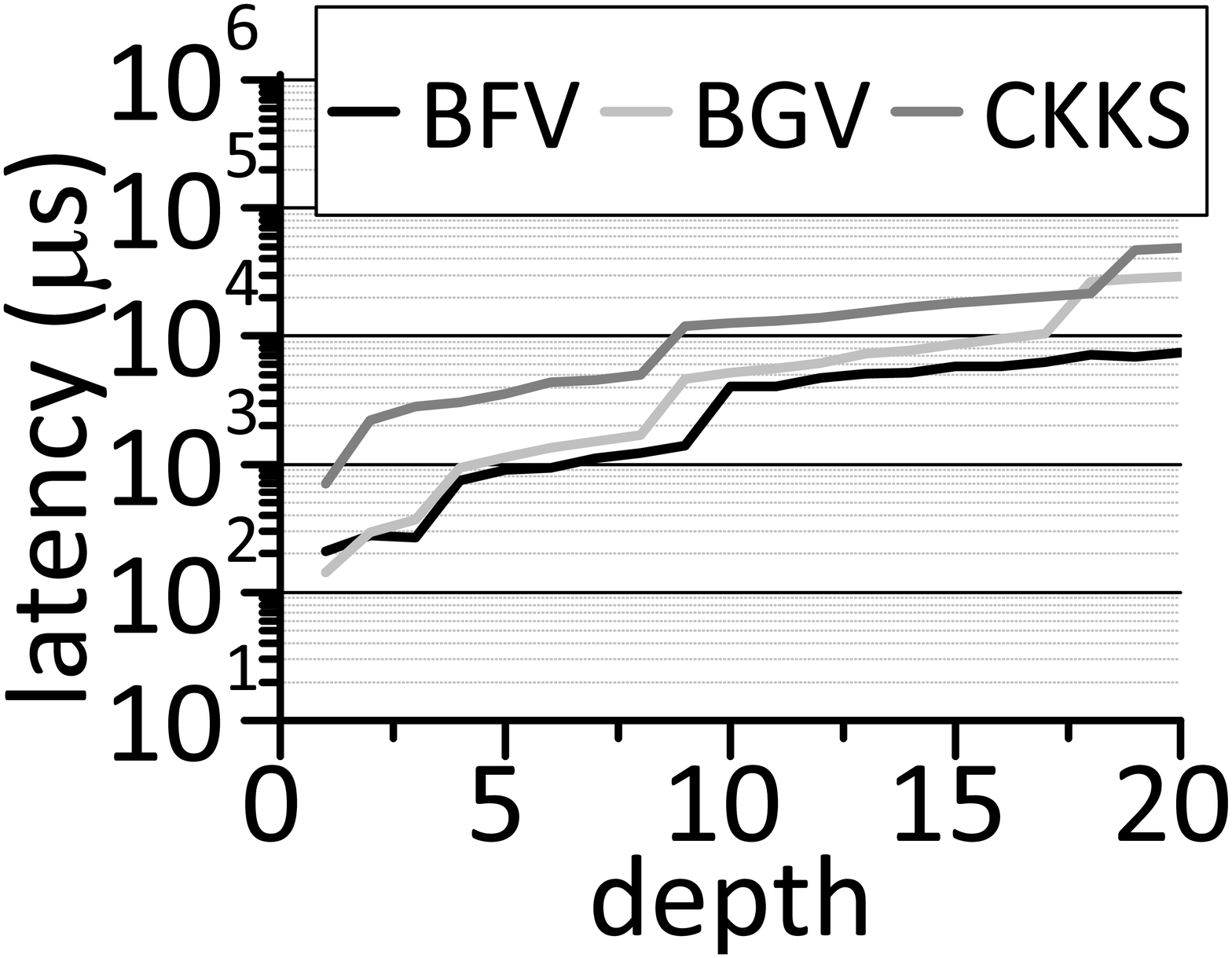}
   \label{f:ben_he_mulcp}
}
\hspace{-0.15in}
\subfigure[Rotation]{
   \includegraphics[width=1.3in]{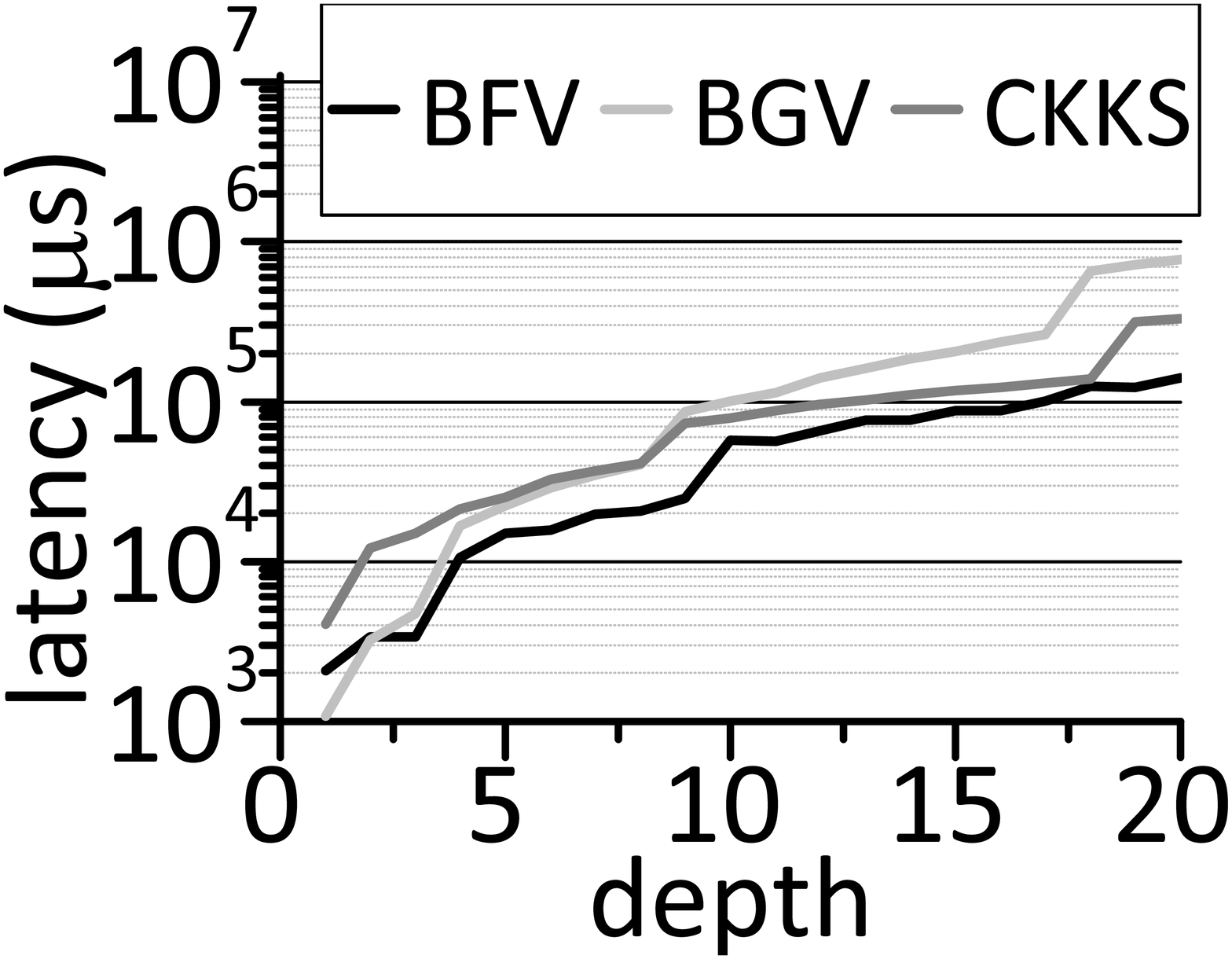}
   \label{f:hen_he_rotation}
}
\vspace{-0.1in}
\caption{The latency comparison of server FHE operations between various schemes.}
\label{f:ben_oper_comparison}
\end{figure*}

\begin{figure*}[t!]
\vspace{-0.1in}
\centering
\subfigure[AddCC]{
   \includegraphics[width=1.3in]{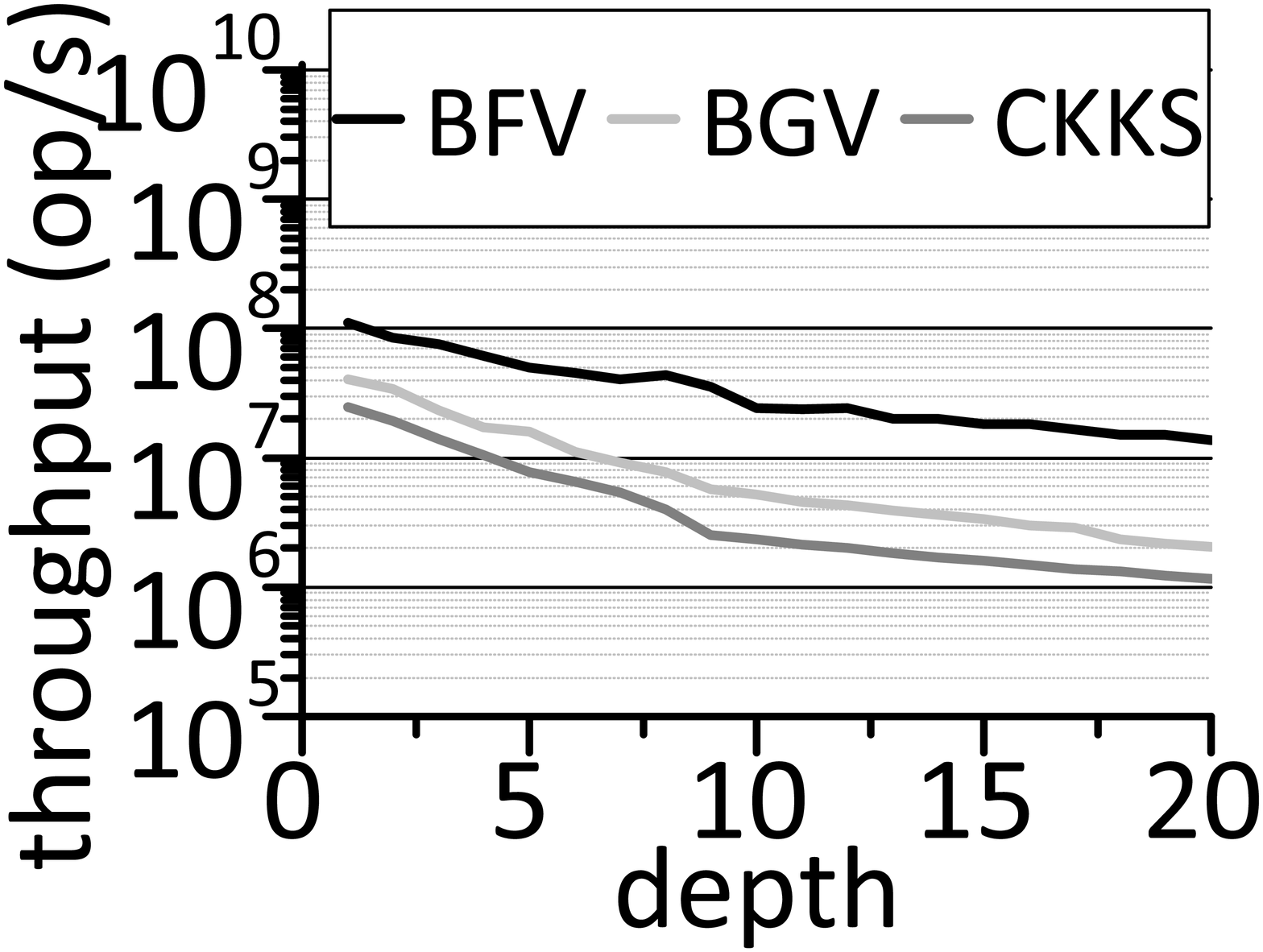}
   \label{f:ben_heavg_addcc}
}
\hspace{-0.15in}
\subfigure[AddCP]{
   \includegraphics[width=1.3in]{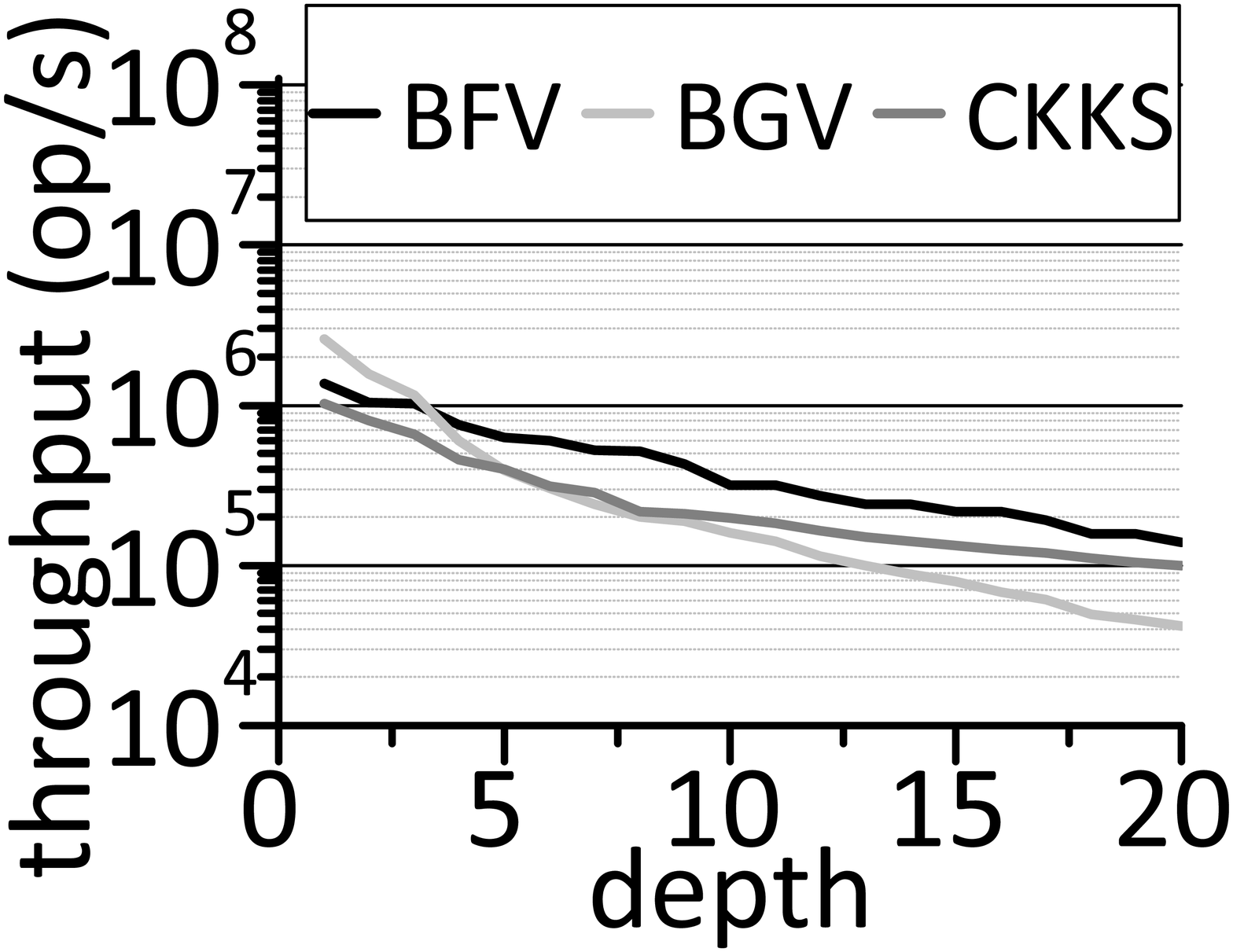}
   \label{f:ben_heavg_addcp}
}
\hspace{-0.15in}
\subfigure[MultCC]{
   \includegraphics[width=1.3in]{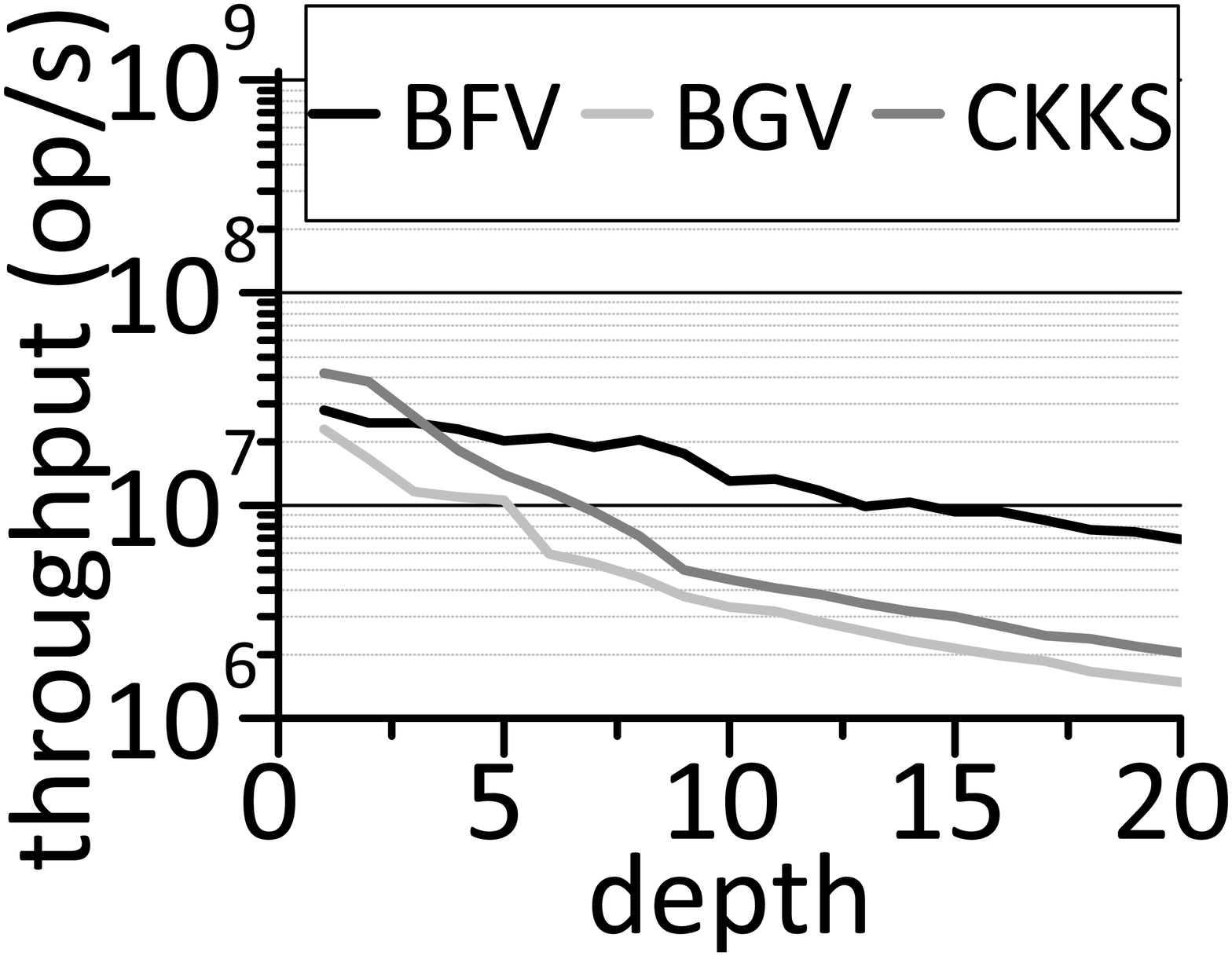}
   \label{f:ben_heavg_mulcc}
}
\hspace{-0.15in}
\subfigure[MultCP]{
   \includegraphics[width=1.3in]{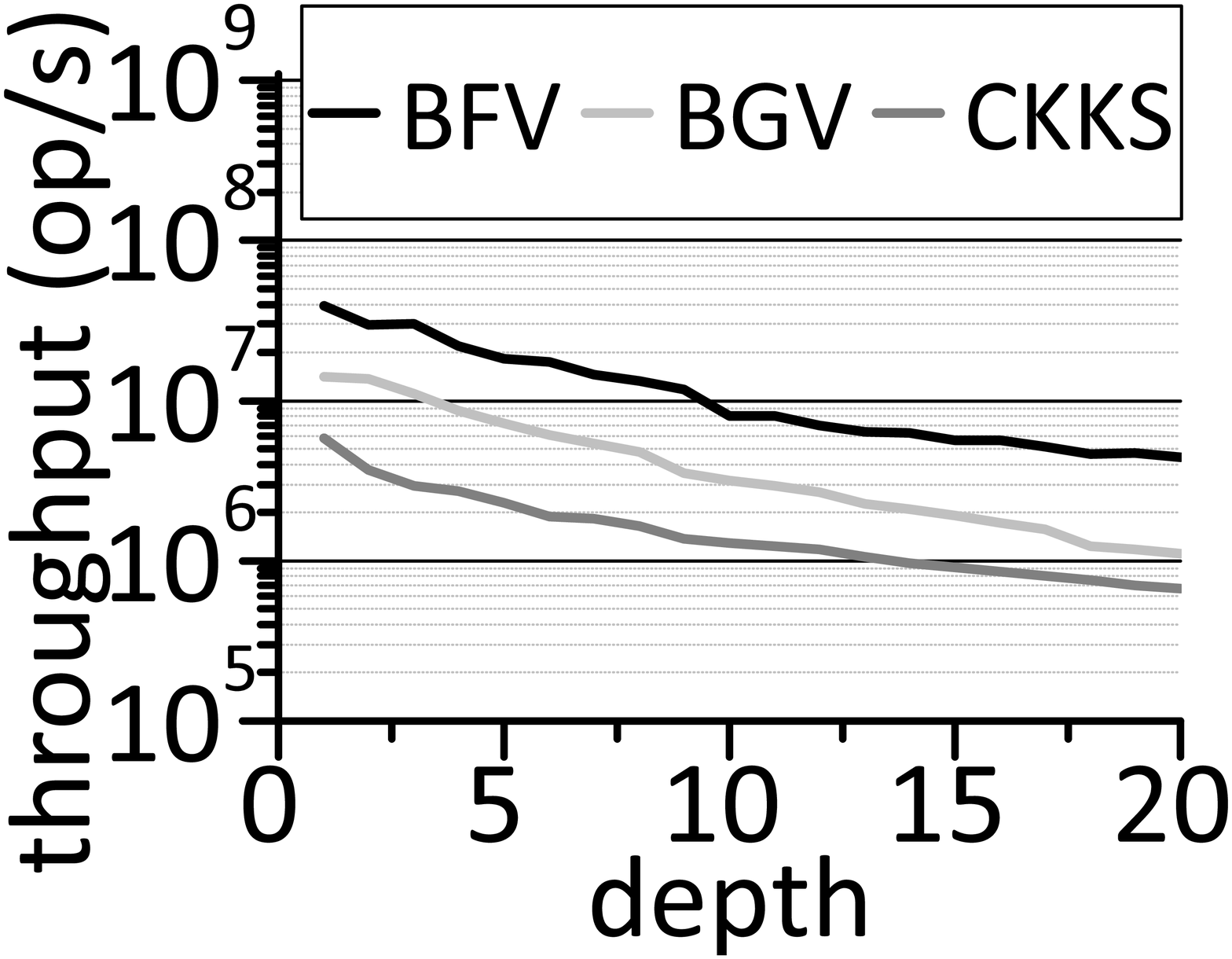}
   \label{f:ben_heavg_mulcp}
}
\hspace{-0.15in}
\subfigure[Rotation]{
   \includegraphics[width=1.3in]{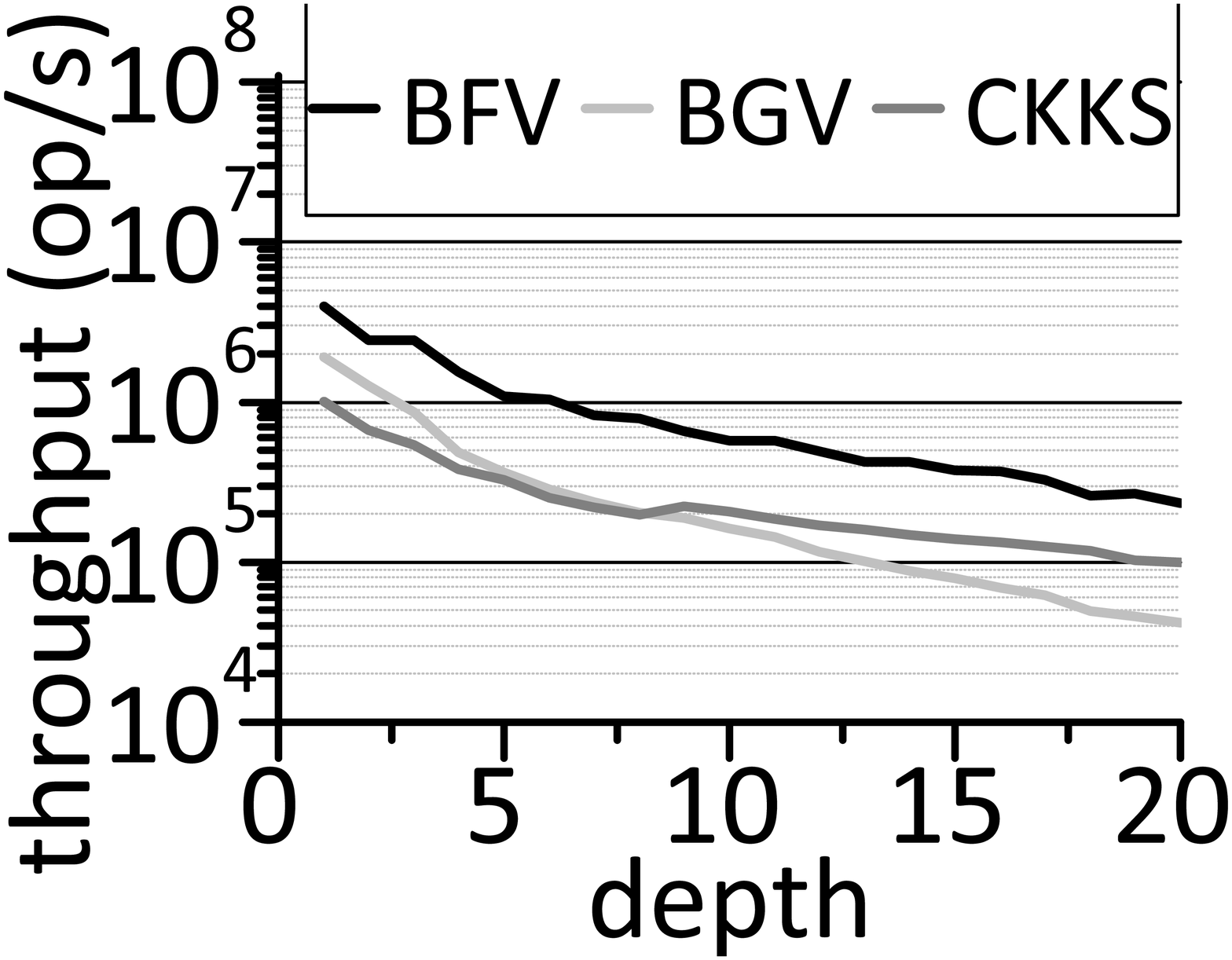}
   \label{f:ben_heavg_rotation}
}
\vspace{-0.1in}
\caption{The throughput comparison of server FHE operations between various schemes.}
\label{f:ben_operavg_comparison}
\vspace{-0.2in}
\end{figure*}

\section{Experimental Methodology}
\label{s:exp_step}

As Table~\ref{t:he_lib_comp} shows, we studied state-of-the-art FHE libraries that implement various FHE schemes. We ran all experiments on our CPU baseline consisting of an Intel Xeon Gold 6138P CPU (20-core, 2GHz) and 64GB DRAM. All libraries are compiled with the Intel icc compiler using the -O3 optimization flag. We used Intel VTune to profile the CPU. Besides key generations, encryption, decryption, and rotation, we evaluated FHE arithmetic operations including an addition between two ciphertexts (\textsf{AddCC}), an addition between a ciphertext and a plaintext (\textsf{AddCP}), a multiplication between two ciphertexts (\textsf{MultCC}), and a multiplication between a ciphertext and a plaintext. For FHE Boolean logic operations, we compared only NAND, AND, and XOR, since the other types of gates share similar performance.

\section{Evaluation and Results}
\label{s:eva_res}

\subsection{FHE Library Comparison}

It is difficult to tell which is the ``best'' FHE library, since no library can attain the strongest security and the fastest speed for all FHE operations. Different libraries have totally different implementation details, e.g., noise estimation, or rescaling method. Our \textit{aim} is to find a competitive library for each FHE scheme to study its performance.

\begin{figure}[t!]
\centering
\subfigure[Arithmetic]{
\includegraphics[width=1.65in]{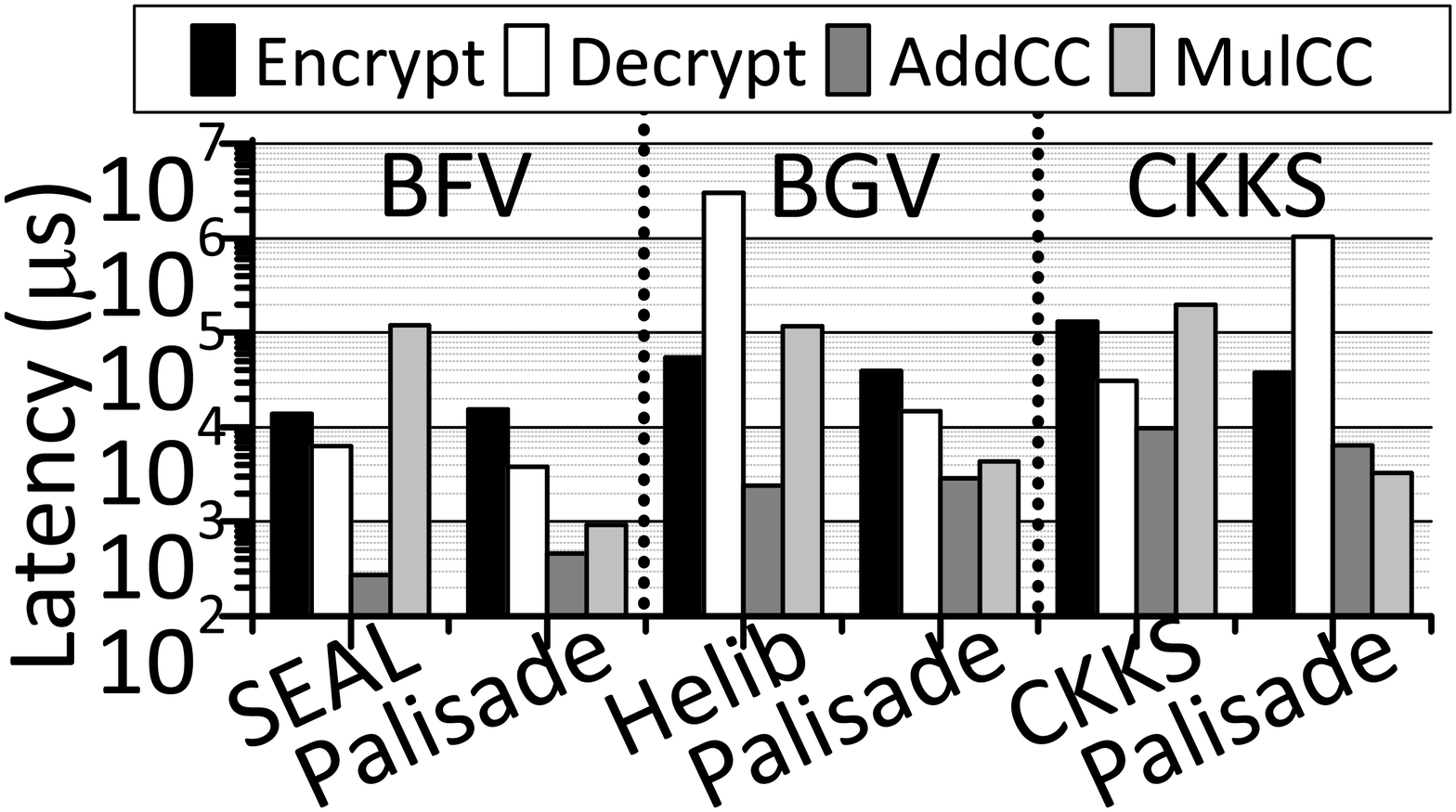}
\label{f:ben_he_liba}

}
\hspace{-0.2in}
\subfigure[Logic]{
\includegraphics[width=1.65in]{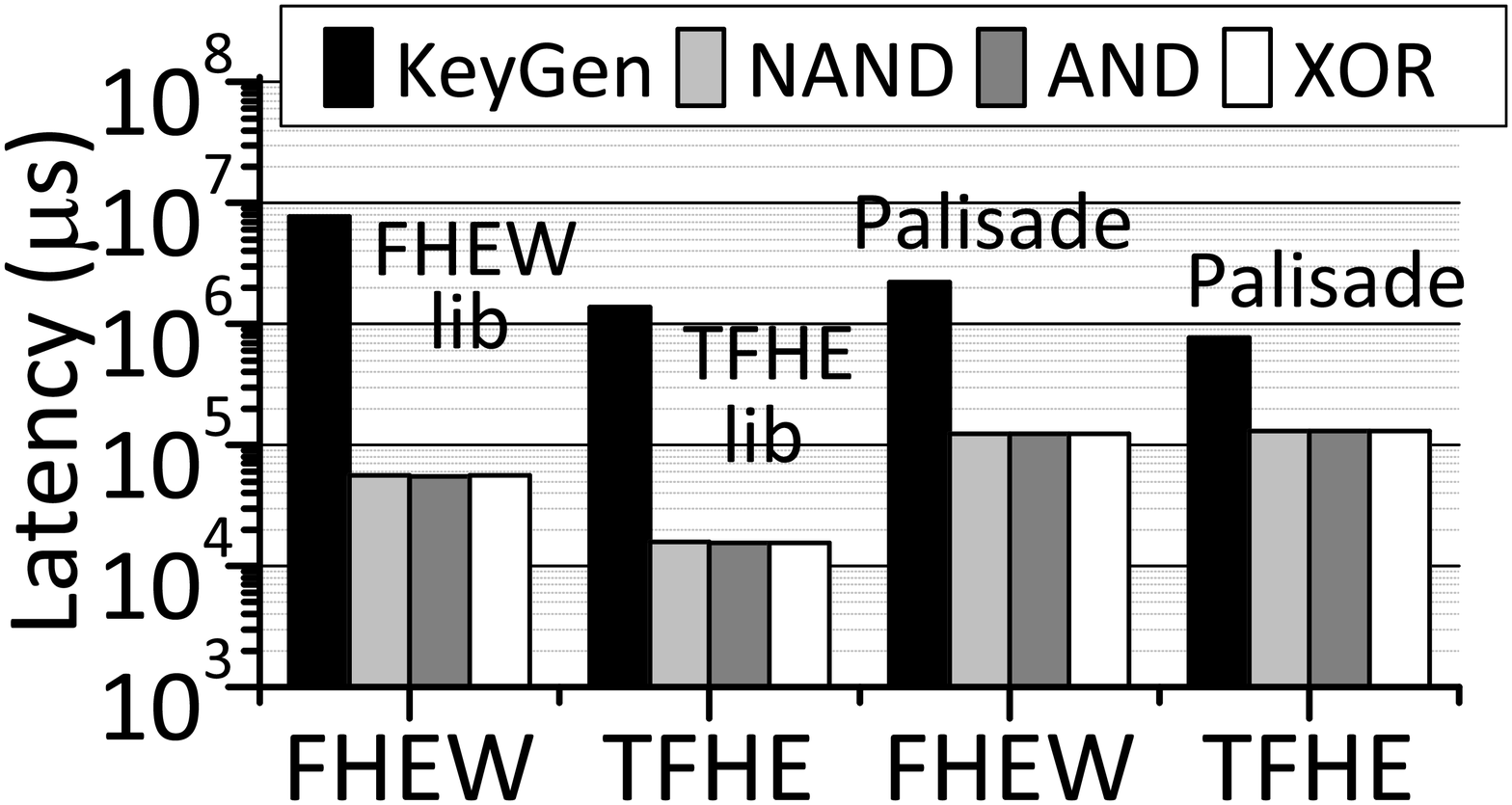}
\label{f:ben_he_libl}
}
\vspace{-0.15in}
\caption{The comparison between FHE libraries (For BFV, $p=786433$, $log_2(n)=14$ and $q=420$. For BGV, $p=786433$, $log_2(n)=15$ and $q=480$. For CKKS, $log_2(p)=32$, $log_2(n)=15$ and $q=350$. For FHEW and TFHE, all operations have boostrapping.).}
\label{f:ben_lib_comparison}
\vspace{-0.25in}
\end{figure}

\textbf{Arithmetic Operation}. The performance comparison of arithmetic operations between various FHE libraries is shown in Figure~\ref{f:ben_he_liba}. Since SEAL and Palisade support no bootstrapping, we consider only leveled FHE, which can support only a certain FHE depth defined as the number of consecutive MultCCs on the critical path without bootstrapping. We fixed the FHE depth as 9, and then derived different values of $p$, $q$ and $n$ for BFV, BGV, and CKKS to guarantee the classical 128-bit security, because we find FHE arithmetic operations with larger depths share a similar trend. No library attains the shortest latencies across all types of arithmetic operations. However, as Figure~\ref{f:ben_he_liba} shows, Palisade achieves the state-of-the-art performance of FHE arithmetic operations across most schemes. More importantly, in Palisade, all schemes share the same data structures and FHE depth estimation algorithm. \textit{So we use Palisade to study the performance of FHE arithmetic operations in the rest of this paper}.

\textbf{Logic Operation}. Figure~\ref{f:ben_he_libl} describes the performance comparison of Boolean logic operations between various FHE libraries. We consider only FHE with bootstrapping for logic operations. We set FHE parameters for FHEW and TFHE libraries to guarantee the 80-bit security in the quantum model~\cite{Chillotti:IACR2018}, which is equivalent to the classical 128-bit security. As Figure~\ref{f:ben_he_liba} describes, the libraries of FHEW and TFHE obtain faster logic operation speed than Palisade. Instead of complicated NTT, FHEW and TFHE libraries use only simple FFT, where there is no modular operation, to accelerate polynomial multiplications~\cite{Chillotti:IACR2018}. However, Palisade still adopts NTT and modular multiplications to implement FHEW and TFHE schemes. Moreover, Palisade does not use X86 SIMD instructions such as AVX512~\cite{Micciancio:IACR2020}. \textit{We select FHEW and TFHE libraries to study the performance of logic operations in the rest of this paper}.

\subsection{FHE Operation Comparison}

\textbf{Arithmetic Operation}. To study arithmetic operations, we chose Palisade. We classify FHE operations into two groups: client and server. The client needs to generate private, public, relinearization, and rotation keys to enable the FHE-based cloud computing. Moreover, the client also has to encrypt and decrypt the sensitive data. On the contrary, the server performs only FHE operations.
\begin{itemize}[nosep,leftmargin=*]
\item \textbf{Client}. The latency comparison of the setup for arithmetic operations on the client side is shown in Figure~\ref{f:ben_peri_comparison}, where a larger FHE depth exponentially increases the setup latency of arithmetic operations. Except the decryption, BGV requires the longest latency when the depth is large, while CKKS has the longest decryption latency. When the FHE depth is small, i.e., $\leq 2$, BFV requires longer latencies of key generation and decryption than BGV. If the target application requires complex or fixed-point arithmetic operations, the client prefers CKKS. For an integer application, when the FHE depth is $>2$, the client favors BFV; otherwise the client may use BGV.

\item \textbf{Server}. The latency comparison of arithmetic operations on the server side is shown in Figure~\ref{f:ben_oper_comparison}, where a larger FHE depth exponentially increases the latency of arithmetic operations. To accommodate a large depth, BGV, BFV and CKKS have to use a larger ciphertext modulus $q$, which greatly prolongs the latency of multiplications and additions. Except the depth of $\leq 2$, BFV uses the shortest latencies for multiplications and additions. For the depth of $\leq2$, BGV has the shortest latencies of multiplications and additions. \textit{With the same FHE depth, AddCP requires longer latency than even MultCC for all three FHE schemes}, since it invokes more frequently NTT kernels. Because BGV, BFV and CKKS support SIMD, we show the throughput of arithmetic operations (operations per second) implemented by three schemes in Figure~\ref{f:ben_operavg_comparison}. BFV achieves the largest throughput for most operations, due to its large batch sizes. If the target application requires complex or fixed-point arithmetic operations, the server prefers CKKS. For an integer application, when the depth is $>2$, the server favors BFV; otherwise the server can use BGV. When maximizing the throughput of \textsf{AddCC} and \textsf{MultCP}, the server should adopt BFV regardless of the depth.
\end{itemize}

\textbf{Logic Operation}. To study FHE logic operations, we ran both FHEW and TFHE libraries. On the client side, since the encryption/decryption latency of FHEW and TFHE is $<5ms$, we evaluate only bootstrapping key generation.
\begin{itemize}[nosep,leftmargin=*]
\item \textbf{Client}. The latency comparison of FHEW and TFHE key generation on the client side is shown in Figure~\ref{f:ben_he_libl}. Compared to FHEW relying on a ternary secret distribution, TFHE uses only a binary secret distribution, which is not compatible with the Homomorphic Encryption Standard~\cite{Micciancio:IACR2020} yet. If the client views TFHE is secure enough, she prefers TFHE due to the shorter key generation latency. 

\item \textbf{Server}. The latency comparison of logic operations on the server side is shown in Figure~\ref{f:ben_he_libl}. The TFHE scheme implemented by its own library is faster, due to its highly vectorized FFT kernel. The TFHE library supports AVX-512 instructions. If the server believes TFHE is secure enough, it prefers TFHE simply because of the speed. Otherwise, the server prefers to use FHEW.
\end{itemize}

\section{Conclusion and Future Work}
\label{s:con}

In this paper, we propose a quantitative study, FHEBench, to study major FHE schemes. We empirically compare state-of-the-art FHE libraries. And then, we choose Palisade, FHEW, and TFHE libraries to study each FHE scheme. Finally, we quantitatively compare the performance of various FHE schemes. FHEBench enables average users to choose the most efficient FHE scheme for their privacy-preserving applications. In future, we will study the performance of FHE schemes on GPUs, FPGAs, and even ASIC-based accelerators.

{\scriptsize
\bibliographystyle{short}
\bibliography{he}
}

\end{document}